\documentclass[10pt,notitlepage,a4paper,aps,prd,onecolumn,superscriptaddress,nofootinbib,groupedaddress]{revtex4-1}
\usepackage{array}

\usepackage[dvipsnames]{xcolor}

\usepackage[pdfencoding=auto, psdextra]{hyperref}
\usepackage[T1]{fontenc}
\usepackage[utf8]{inputenc}
\usepackage[english]{babel}
\usepackage{hyperref}
\usepackage[mathscr]{euscript}
\usepackage{caption}
\usepackage{subcaption}
\hypersetup{
colorlinks = true,
linkcolor = blue,
filecolor = magenta,
urlcolor = blue,
pdftitle = {CP3 entanglement},
pdfpagemode = FullScreen,
citecolor = blue}
\usepackage{physics}

\usepackage{caption}

\usepackage{indentfirst}

\usepackage[pdftex]{graphicx}

\usepackage{amsmath}
\allowdisplaybreaks  
\usepackage{amssymb} 
\usepackage{amsfonts} 
\usepackage{enumerate}

\usepackage{fancyhdr}                     

\usepackage{float}                             

\usepackage{amsmath}         
\usepackage{amsfonts}        
\usepackage{amssymb}         
\usepackage{amsthm}          
\usepackage{empheq}          

\usepackage{color}           
\usepackage{ragged2e}        
\usepackage{titlesec}        

\usepackage{tipa}            

\usepackage[utf8]{inputenc} 
\allowdisplaybreaks          

\usepackage{enumitem}        

\usepackage{tikz}            
\usetikzlibrary{shapes.geometric,arrows.meta,arrows,positioning}  
\usetikzlibrary{tikzmark}   

\titleformat{\paragraph}
  {\normalfont\normalsize\bfseries}{\theparagraph}{1em}{} 
\titlespacing*{\paragraph}
  {0pt}{3.25ex plus 1ex minus .2ex}{1.5ex plus .2ex} 

\begin{document}
\title{An Exact Five-Step Method for Classicalizing N-level Quantum Systems: Application to Quantum Entanglement Dynamics}

\author{Daniel Martínez-Gil}
\email{daniel.martinez@ua.es}
\affiliation{Fundacion Humanismo y Ciencia, Guzmán el Bueno, 66, 28015 Madrid, Spain.}
\affiliation{Departamento de F\'{\i}sica Aplicada, Universidad de Alicante, Campus de San Vicente del Raspeig, E-03690 Alicante, Spain.}

\author{Pedro Bargueño}
\email{pedro.bargueno@ua.es}
\affiliation{Departamento de F\'{\i}sica Aplicada, Universidad de Alicante, Campus de San Vicente del Raspeig, E-03690 Alicante, Spain.}

\author{Salvador Miret-Artés}
\email{s.miret@iff.csic.es}
\affiliation{Instituto de Física Fundamental, Consejo Superior de Investigaciones Científicas, Serrano 123, 28006, Madrid, Spain}

\begin{abstract}
    In this manuscript, we present a general and exact method for classicalizing the dynamics of any $N$-level quantum system, transforming quantum evolution into a classical-like framework using the geometry of complex projective spaces $\mathbb{CP}^{N-1}$. The method can be expressed as five-step algorithmic procedure to derive a classical Hamiltonian and a symplectic structure for the Poisson brackets, yielding $N-1$ Hamilton’s equations that precisely replicate the quantum dynamics, including complex phenomena like entanglement. We demonstrate the method’s efficacy by classicalizing two interacting qubits in $\mathbb{CP}^3$, exactly reproducing quantum observables such as quantum probabilities, quaternionic population differences and the concurrence, capturing entanglement dynamics via a classical analog. 
\end{abstract}

\maketitle

\section{Introduction}
The interplay between quantum and classical mechanics has been a long key topic of modern science, influencing different researches across multiple disciplines.
 Quantum mechanics, 
 with its foundation in complex Hilbert spaces and non-commuting operators, offers an extraordinary precision in describing microscopic phenomena, from the behavior of subatomic particles to the properties of molecules and materials. 
 However, this precision 
has a considerable cost:
 as the number of degrees of freedom in a system grows, the dimensionality of the Hilbert space increases exponentially, making the direct quantum simulations highly computationally demanding for all but the simplest cases. This challenge becomes especially pronounced in many-body systems, where the exponential growth of the quantum state space overwhelms even the most advanced computational resources. Classical mechanics, by contrast, operates in the intuitive realm of phase space trajectories and Poisson brackets, providing a computationally tractable framework that aligns naturally with our macroscopic experience. 
 Therefore, the topic of constructing classical analogues that replicate quantum dynamics (exactly or with high accuracy) emerged as a new strategy, unlocking new ways of exploring across physics and chemistry.

The motivation for classicalization is based on both practical and conceptual considerations. For example, quantum phenomena such as coherence, tunneling, and entanglement are not mere theoretical abstractions; they govern the behavior of quantum systems and technologies, which play a central role in condensed matter \cite{AltlandSimons2010,Sachdev2011}, many-body physics \cite{Bloch2008}, or quantum information \cite{NielsenChuang2010}, or molecular reactions \cite{SzaboOstlund1996}, among others.  As these systems grow the dimension, their exact quantum study becomes difficult; their classical framework might offering an attractive alternative, capturing the essence of quantum dynamics into a form that is both computationally efficient and conceptually attractive.

Historically, efforts to classicalize quantum mechanics have yielded valuable insights. An early work by Strocchi \cite{strocchi} rewrote the Schrödinger equation into a Hamiltonian framework, using the expectation value of the quantum Hamiltonian as a classical function. This approach revealed that quantum evolution could, in principle, be mirrored by classical-like dynamics, serving as a starting point for subsequent developments. 
In molecular physics, Meyer and Miller \cite{Meyermiller} 
pioneered a mapping for electronic Hamiltonians, later refined by Stock and Thoss \cite{StockThoss}, which has proven for simulating electronic transitions \cite{Tully1990}. More recent innovations, including Cotton and Miller’s spin-mapping techniques \cite{CottonMiller} and the Stratonovich–Weyl formalism of Runeson and Richardson \cite{Richardson1,Richardson2}, have extended these ideas to multilevel systems, often under quasiclassical approximations.
On the other hand, geometric perspectives of classicalization can be performed, which were first formulated by Kibble \cite{Kibble}. Subsequent advances were performed 
 by Gibbons \cite{GIBBONS1992147}, Ashtekar and Schilling \cite{Ashtekar1999}, and Brody and Hughston \cite{BRODY200119} among others, introducing complex projective spaces as a natural way for quantum-classical correspondence, offering mathematical elegance and physical intuition. 

Despite these advances, significant challenges remain. Many existing methods are specific for the system under study, or are based on approximations that fail in some regimes. These limitations justify the need for a general, exact classicalization framework capable of transcending disciplinary boundaries.

In this paper, we explicitly implement a \emph{general and exact} method for classicalizing any $N$-level quantum system, applicable across the spectrum of physics and chemistry. The method uses a geometric representation where the normalized quantum state corresponds to a point in a well-defined phase space. 
At this point, we emphasize that, while the mathematical existence of a classical counterpart for any $N$-level quantum system in $\mathbb{CP}^{N-1}$ is well-established \cite{book_geometry_quantum}, we have no found previous work which explicitly formulated and applied a systematic method to realize this correspondence. In our work, through a systematic five-step algorithm, we derive a classical Hamiltonian function and its associated dynamical structure, yielding a system of $N-1$ Hamilton’s equations. These equations faithfully reproduce the quantum evolution of the original system, capturing complex phenomena such as entanglement without approximation. 
The exact correspondence not only gives a geometric interpretation through which to interpret the quantum behaviour, but also provides an enhancement of computational efficiency, reducing the number of solving equations and working with funtions instead of with matrices.

To demonstrate the efficacy of the method, we apply it to a paradigmatic system: two interacting qubits. This case is particularly instructive, as entanglement plays a central role in quantum information processing, from cryptography to teleportation \cite{NielsenChuang2010}. The classical representation explained in this work, exactly reproduces the quantum dynamics, including state populations, quaternionic differences, and concurrence, even under entangling interactions. This example
 highlights the potential of the method, offering a relationship between quantum complexity and classical simplicity.
 
Specifically, this work is organized as follows. In Section \ref{method}, we detail the geometric framework and the five-step algorithm for classicalization. In Section \ref{entanglement}, we propose a two-qubit interaction example, showing its exact classical correspondence. In Section \ref{results}, we show the results comparing the quantum and classical frameworks. Finally, Section \ref{conclusions} provides the conclusions of the present work.

\section{Method}\label{method}

For a simple two-state quantum system, such as an isolated qubit, classicalization can be achieved intuitively. By taking the expectation value of the Hamiltonian and employing canonical Hamilton's equations, the quantum dynamics are mapped to a classical phase space. This method transforms the quantum problem into a set of classical differential equations. However, for more complex systems (as two entangled qubits) this process 
 requires adjustments to capture the entanglement between states.

A notable result from quantum geometry extends this idea: any quantum system with $N$ states can be exactly mapped to a classical system with $N-1$ states \cite{book_geometry_quantum}.
 Within this framework, quantum trajectories are reinterpreted as solutions to Hamilton's equations. Using complex canonical coordinates, the Schrödinger equation can be reformulated as a classical system, where the Hamiltonian function is the average value of the quantum Hamiltonian operator in these coordinates. This connection suggests that quantum evolution can be understood as a special case of classical dynamics in an appropriately defined phase space.

For any quantum system with $N$ states, the state space is the complex projective space $\mathbb{CP}^{N-1}$ \cite{book_geometry_quantum}. This space manifests the irrelevance of global phases in quantum mechanics and provides a geometric alternative for studying quantum dynamics. Importantly, $\mathbb{CP}^{N-1}$ is a Kähler manifold \cite{Moroianu_2007, Bellmann, Nakahara}, equipped with the Fubini-Study metric (defining distances)\cite{fubini, study}, a symplectic form (enabling Hamiltonian dynamics)\cite{arnold, Lee}, and a complex structure \cite{chern-1979, Griffiths, harris-1992}. Symplectic forms are well-known in a classical dynamics context, being essential for defining Hamilton's equations on manifolds, providing the geometric structure for Hamiltonian mechanics \cite{Arnold1989}. A symplectic form is a closed, non-degenerate 2-form that enables the definition of the Poisson bracket, a key operation in dynamics, as we will see.

The Fubini-Study metric and the symplectic form can be related to each other in $\mathbb{CP}^{N-1}$
 maintaining a structure consistent with Kähler spaces.
 This consistency ensures that the geometric properties scale predictably with dimension, facilitating the transition from quantum to classical-like descriptions. This consistency also enables us to  exactly classicalize any $N$-state quantum system by considering
 its evolution as a classical Hamiltonian flow within $\mathbb{CP}^{N-1}$, which is the primary strenght of this model.

In the following, we will specify a five-step method to classicalize a N-state quantum system.

\subsection*{Five steps to classicalize a N-state quantum system}

Given a quantum Hamiltonian, the process of ``classicalizing'' it into a suitable form on $\mathbb{CP}^{N-1}$ consists of five automatable steps. These steps systematically transform the quantum system into a classical-like framework while preserving its essential physical behavior.

\subsubsection*{1. Obtain coordinates in $\mathbb{CP}^{N-1}$}

We begin with a wave function for a system of $N$ states, expressed in a chosen basis as

\begin{equation}
\ket{\psi} = \sum_{i=0}^{N-1} a^i \ket{\phi_i},
\end{equation}

where $a^i$ are complex coefficients and $\ket{\phi_i}$ are the basis states. To define coordinates in $\mathbb{CP}^{N-1}$, we select a non-zero coefficient, say $a^{N-1}$, and compute the inhomogeneous coordinates:

\begin{equation}
x^i = \frac{a^i}{a^{N-1}} \quad \text{for} \quad i = 0, 1, \ldots, N-2.
\end{equation}

This yields $N-1$ independent complex coordinates, parameterizing the state in $\mathbb{CP}^{N-1}$.

\subsubsection*{2. Express the wave function in $\mathbb{CP}^{N-1}$ coordinates}

Using these coordinates, the wave function can be rewritten as

\begin{equation}
\ket{\psi} = \frac{1}{\sqrt{\mathcal{N}}} \left( \sum_{i=0}^{N-2} x^i \ket{\phi_i} + \ket{\phi_{N-1}} \right),
\end{equation}

where $\mathcal{N} = 1 + \sum_{i=0}^{N-2} |x^i|^2$ is the normalization factor in this representation. Since $\mathbb{CP}^{N-1}$ is a projective space, the overall factor $a^{N-1}$ does not affect the state's physical properties, and the dynamics are determined by the coordinates $x^i$.

\subsubsection*{3. Compute the Hamiltonian function}

The classical-like Hamiltonian $H_0$ is obtained by calculating the expectation value of the quantum Hamiltonian $\hat{H}$ with respect to $\ket{\psi}$, as

\begin{equation}
H_0 = \langle \psi | \hat{H} | \psi \rangle.
\end{equation}

This scalar function $H_0$ represents the Hamiltonian governing classical-like dynamics on $\mathbb{CP}^{N-1}$.

\subsubsection*{4. Determine the symplectic form of the space}

In order to determine the symplectic form of $\mathbb{CP}^{N-1}$, a quantity $K$, called Kähler potential, is defined in terms of the normalization factor as follows

\begin{equation}
K = \log \left( \mathcal{N} \right).
\end{equation}

From this potential, the Fubini-Study metric $g_{j \bar{k}}$ can be derived as

\begin{equation}
g_{j \bar{k}} = 2 \frac{\partial^2 K}{\partial x^j \partial \bar{x}^k}.
\end{equation}

Having defined the metric of the space, we can relate it to the symplectic form matrix as

\begin{equation}
\omega_{j \bar{k}} = \frac{i}{2} g_{j \bar{k}} = i \left( \frac{\delta_{j k}}{\mathcal{N}} - \frac{x^j \bar{x}^k}{\mathcal{N}^2} \right),
\end{equation}
with its inverse being
\begin{equation}
\omega^{j \bar{k}} = -i \mathcal{N} \left( \delta^{j k} + x^j \bar{x}^k \right),
\end{equation}
where indices $j, k$ run from 0 to $N-2$. 

\subsubsection*{5. Derive Hamilton's Equations}

Finally, the classical dynamics are described by Hamilton's equations, which can be generally written in terms of Poisson brackets as follows
\begin{equation}
    \dot{x}^j = \{x^j, H_0\},
\end{equation}
where the Poisson bracket can be defined in terms of the inverse of the symplectic form matrix as
\begin{equation}\label{poisson}
    \{f,H\} = \sum_{j,k = 0}^{N-2}\omega^{jk} \left(\frac{\partial f}{\partial q^j}\frac{\partial H}{\partial p^k}-\frac{\partial f}{\partial p^j}\frac{\partial H}{\partial q^k}\right),
\end{equation}
being $q, p$ generalized position and momentum. Note that, in flat spaces such as $\mathbb{R}^{2n}$, the canonical symplectic form is used, $\omega^{jk} = 1$, yielding the usual Hamilton's equations $\left(\frac{dq_i}{dt} = \frac{\partial H}{\partial p_i}, \quad \frac{dp_i}{dt} = -\frac{\partial H}{\partial q_i}\right)$. In our case, considering $(x^j, \bar{x}^j)$ as a pair of canonically conjugate variables, we can obtain the general expression for Hamilton's equations in $\mathbb{CP}^{N-1}$ as
\begin{equation}
\dot{x}^j = \sum_{k=0}^{N-2} -i \mathcal{N} (\delta^{j k} + x^j \bar{x}^k)\frac{\partial H_0}{\partial \bar{x}^k}.
\end{equation}

This formulation reduces the original $N$ quantum equations to $N-1$ classical equations, enabling the simulation of quantum behavior in a lower-dimensional classical framework.

\section{Entanglement mapping}\label{entanglement}

Having outlined the general approach to classicalize quantum systems, we can now focus on a specific model: a system of two entangled qubits, the entanglement being a well-known purely quantum quantity. 

To classicalize the system, we first require a consistent quantum model. The total wave function of two entangled qubits can be expressed as
\begin{equation}
    \ket{\psi} = a \ket{00}+b\ket{10}+c\ket{01}+d\ket{11},
\end{equation}
where $|a|^2 + |b|^2 + |c|^2 + |d|^2 = 1$. Since $a, b, c, d$ are complex coefficients, $\ket{\psi}$ resides in $S^7$.

To evolve this wave function over time via the Schrödinger equation, we need a Hamiltonian operator $\hat{H} \in \mathrm{SU}(4)$. Constructing a basis for $\mathrm{SU}(4)$ requires 15 elements, which, in terms of Pauli matrices, it can be written as
\begin{equation}
    \text{Basis}(SU(4)) = \left\{\hat{\sigma}_i \otimes \hat{\sigma}_j | i, j = 0,x,y,z \right\},
\end{equation}
where $\hat{\sigma}_0 \otimes \hat{\sigma}_0$ is excluded to yield 15 elements, and $\hat{\sigma}_0 = \hat{I}_{2 \times 2}$ represents the $2 \times 2$ identity matrix. Notably, only terms with $\hat{\sigma}_i$ and $\hat{\sigma}_j \neq \hat{I}$ contribute to the entanglement.

In order to study an entangled system, we consider the following example Hamiltonian:
\begin{equation}\label{Hamcuantico}
    \hat{H} = C_1 (\hat{\sigma}_z \otimes \hat{I})+C_2(\hat{\sigma}_x \otimes \hat{I}) +C_3 (\hat{\sigma}_y \otimes \hat{I}) + C_4 (\hat{\sigma}_y \otimes \hat{\sigma}_y)+C_5 (\hat{\sigma}_x \otimes \hat{\sigma}_y),
\end{equation}
where the first three terms represent the standard Hamiltonian in the $\mathrm{SU}(4)$ basis, and the last two terms encode the system’s entanglement. Applying the time-dependent Schrödinger equation
\begin{equation}
    i\hbar\frac{d}{dt} \ket{\psi} = \hat{H}\ket{\psi},
\end{equation}
and considering $\hbar = 1$, we obtain
\begin{eqnarray}
    i\dot{a} & = C_1 a + (C_2-iC_3) c + (-C_4-iC_5)d, \\
    i\dot{b} & = C_1 b + (C_2-iC_3) d + (C_4+iC_5)c,\\
    i\dot{c} & = -C_1 c + (C_2+iC_3) a + (C_4-iC_5)b,\\
    i\dot{d} & = -C_1 d + (C_2+iC_3) b + (-C_4+iC_5)a.
\end{eqnarray}

To quantify the entanglement of the system, we use the concurrence, defined by Wootters \cite{wootters}. For our system, the concurrence is given by $C = 2 |ad - bc|$ \cite{Mosseri_2001}, where $C = 1$ indicates a maximally entangled state, and $C = 0$ corresponds to a non-entangled state. 
Once we have introduced the quantum model, we can start to classicalize it.

\subsection{Classical model}
 
Interestingly, in a previous work \cite{dani2}, we examined the classicalization of a two-level system (a single qubit), where the state space is
$\mathbb{CP}^1$. We want to remark on that $\mathbb{CP}^1$ is isomorphic to the two-sphere $S^2$ (widely recognized as the Bloch sphere) and that is why the first Hopf fibration can provide a relationship between classical and quantum mechanics.

While the second Hopf fibration maps the seven-sphere $S^7$ to the four-sphere $S^4$, this construction is not isomorphic to $\mathbb{CP}^3$. Therefore, to construct an exact classical representation of this quantum system, we must work within its natural geometric framework, the complex projective space $\mathbb{CP}^3$, which corresponds to a four-state system. 
 This formalism, although more complex than the single-qubit case, ensures an exact representation of the quantum dynamics as was explained in the previous section, including the effects of entanglement. It should be emphasized that, while the geometry of quantum entanglement has been extensively studied \cite{Mosseri_2001, Levay_2004, Kus, Bengtsson, Bernevig_2003, Kiosses_2014}, an explicit model for exactly classicalizing entanglement has not yet been developed, up to our knowledge.

To classicalize the entangled two-qubit model, we first define coordinates in $\mathbb{CP}^3$. The three coordinates are defined as
\begin{align}
    x^0 &= \frac{a}{d}, &x^1 &= \frac{b}{d},
    &x^2&=\frac{c}{d},
\end{align}
where we assume $d \neq 0$. Note that $x^0, x^1, x^2$ can also be defined by dividing by $a$, $b$, or $c$. Consequently, the wave function can be expressed as
\begin{align}
    \ket{\psi} &= \frac{1}{\sqrt{\mathcal{N}}}(x^0\ket{00}+x^1\ket{01}+x^2\ket{10}+\ket{11}),
\end{align}
where the normalization $\mathcal{N}$ is
\begin{equation}
    \mathcal{N} = 1+\abs{x^0}^2+\abs{x^1}^2+\abs{x^2}^2.
\end{equation}

As $\mathbb{CP}^3$ is a Kähler space, we can define the Kähler potential, metric and symplectic form matrix as in the previous section, finally obtaining the inverse symplectic form matrix as
\begin{equation}\label{inv_symplect}
    \omega^{jk} = -i\mathcal{N}(\delta^{jk} + \omega^i\Bar{\omega}^k).
\end{equation}

Finally, in order to apply Hamilton equations, a Hamiltonian function is needed. This Hamiltonian function can be obtained by performing $ H_0 = \bra{\psi}\hat{H}\ket{\psi}$ in the wave function in $\mathbb{CP}^3$ coordinates, obtaining
\begin{equation}
   H_0 = C_1\bra{\psi}\hat{\sigma}_z \otimes \hat{I})\ket{\psi}+C_2\bra{\psi}\hat{\sigma}_x \otimes \hat{I}\ket{\psi}+ C_3\bra{\psi}\hat{\sigma}_y \otimes \hat{I}\ket{\psi}+C_4\bra{\psi}\hat{\sigma}_y \otimes \hat{\sigma}_y\ket{\psi}+C_5\bra{\psi}\hat{\sigma}_x \otimes \hat{\sigma}_y\ket{\psi},
\end{equation}

where
\begin{align*}
    \bra{\psi}\hat{\sigma}_z \otimes \hat{I}\ket{\psi} &= \frac{C_1}{\mathcal{N}}(\abs{x^0}^2+\abs{x^1}^2-\abs{x^2}^2-1),\\
    \bra{\psi}\hat{\sigma}_x \otimes \hat{I}\ket{\psi} &=\frac{C_2}{\mathcal{N}}(x^2\Bar{x}^0+\Bar{x}^1+\Bar{x}^2x^0+x^1),\\
    \bra{\psi}\hat{\sigma}_y \otimes \hat{I}\ket{\psi} &=i\frac{C_3}{\mathcal{N}}(-x^2\Bar{x}^0-\Bar{x}^1+\Bar{x}^2x^0+x^1),\\
    \bra{\psi}\hat{\sigma}_y \otimes \hat{\sigma}_y\ket{\psi} &= \frac{C_4}{\mathcal{N}}(-\Bar{x}^0+x^2\Bar{x}^1+x^1\Bar{x}^2-x^0),\\
    \bra{\psi}\hat{\sigma}_x \otimes \hat{\sigma}_y\ket{\psi} &= i\frac{C_5}{\mathcal{N}}(-\Bar{x}^0+x^2\Bar{x}^1-x^1\Bar{x}^2+x^0).\\
\end{align*}

Therefore, as was explained in the previous section, considering $x^j, \Bar{x}^j$ as pairs of canonically conjugate variables, and using the Poisson bracket \eqref{poisson} and the inverse symplectic form matrix \eqref{inv_symplect}, we obtain 
\begin{equation}
    \dot{x}^j = \sum_{j,k = 0}^2 -i\mathcal{N}(\delta^{jk}+x^j\Bar{x}^k)\frac{\partial H_0}{\partial \Bar{x}^k}.
\end{equation}

Specifically, the three Hamilton's equations can be expressed as 
\begin{align}
    \dot{x}^0 & = -i\mathcal{N}\left[(1+x^0\Bar{x}^0)\frac{\partial H_0}{\partial \Bar{x}^0}+x^0\Bar{x}^1\frac{\partial H_0}{\partial \Bar{x}^1}+x^0\Bar{x}^2\frac{\partial H_0}{\partial \Bar{x}^2}\right],\\
    \dot{x}^1 & = -i\mathcal{N}\left[x^1\Bar{x}^0\frac{\partial H_0}{\partial \Bar{x}^0}+(1+x^1\Bar{x}^1)\frac{\partial H_0}{\partial \Bar{x}^1}+x^1\Bar{x}^2\frac{\partial H_0}{\partial \Bar{x}^2}\right],\\
    \dot{x}^2 & = -i\mathcal{N}\left[x^2\Bar{x}^0\frac{\partial H_0}{\partial \Bar{x}^0}+x^2\Bar{x}^1\frac{\partial H_0}{\partial \Bar{x}^1}+(1+x^2\Bar{x}^2)\frac{\partial H_0}{\partial \Bar{x}^2}\right],
\end{align}
where, if we write $H_0 = \frac{D}{\mathcal{N}}$,
\begin{align}
    \frac{\partial H_0}{\partial \Bar{x}^0} &=\frac{(C_1 x^0+C_2 x^2-iC_3 x^2-C_4-iC_5)\mathcal{N}-D x^0}{\mathcal{N}^2}, \\
    \frac{\partial H_0}{\partial \Bar{x}^1} &=\frac{(C_1 x^1+C_2 -iC_3 +C_4 x^2+iC_5 x^2)\mathcal{N}-D x^1}{\mathcal{N}^2}, \\
    \frac{\partial H_0}{\partial \Bar{x}^2} &=\frac{(-C_1 x^2 +C_2 x^0+iC_3 x^0+C_4 x^1-iC_5 x^1)\mathcal{N}-D x^2}{\mathcal{N}^2}.
\end{align}

To classicalize the entangled two-qubit model, we have reduced the quantum description from four equations to three in the classical framework, which fully capture the quantum dynamics. This classical formalism also allows us to define an analog to the quantum concurrence, expressed as 
\begin{equation}\label{concurrence}
C = 2\frac{\abs{x^0 - x^1 x^2}}{\mathcal{N}}.
\end{equation}

At this point, some comments are in order. As we previously introduced, the concurrence is a measure of separability, indicating the degree of entanglement in the system. In quantum mechanics, a concurrence of $C = 0$ implies a separable state, where the qubits are independent and not entangled. In the classical formalism, this condition $C = 0$ has a deep geometric interpretation, corresponding to the Segre embedding \cite{book_geometry_quantum}. The Segre embedding maps the product of two projective spaces, $\mathbb{CP}^1 \times \mathbb{CP}^1$, into the higher-dimensional projective space $\mathbb{CP}^3$. When $C = 0$, the system lies on the submanifold $\mathbb{CP}^1 \times \mathbb{CP}^1 \subset \mathbb{CP}^3$, allowing us to study its dynamics within this subspace. Since $\mathbb{CP}^1$ is equivalent to the 2-sphere ($S^2$), we have $\mathbb{CP}^1 \times \mathbb{CP}^1 \approx S^2 \times S^2$, which represents the state space of two non-entangled qubits, each one described independently by $S^2$. Therefore, geometrically, the condition $C=0$ defines a Segree embedding, enabling us to analyze the dynamics of an entangled system in $\mathbb{CP}^3$ or $S^7$ within the non-entangled subspace $S^2 \times S^2$.

\section{Results}\label{results}
In this section, we aim to show that the classical-like formalism exactly reproduces the quantum dynamics, including the entanglement. Given the extensive range of possible results, we will focus on presenting only some relevant ones. For each case, we will provide the result obtained from the Schrödinger equation and its corresponding classical analog, ensuring a clear comparison between the two frameworks.

First of all, we present the quantum probabilities $\abs{a}^2$, $\abs{b}^2$, $\abs{c}^2$, and $\abs{d}^2$ as functions of time, together with their classical counterparts. These classical analogues can be expressed in terms of the coordinates on $\mathbb{CP}^3$ as follows:
\begin{align}
    \abs{a}^2 &= \frac{\abs{x^0}^2}{\mathcal{N}}, &
    \abs{b}^2 &= \frac{\abs{x^1}^2}{\mathcal{N}}, &
    \abs{c}^2 &= \frac{\abs{x^2}^2}{\mathcal{N}}, &
    \abs{d}^2 &= \frac{1}{\mathcal{N}}.
\end{align}

\begin{figure}[H]
	\begin{subfigure}[b]{0.49\textwidth}
		\includegraphics[width=\textwidth, height=7cm]{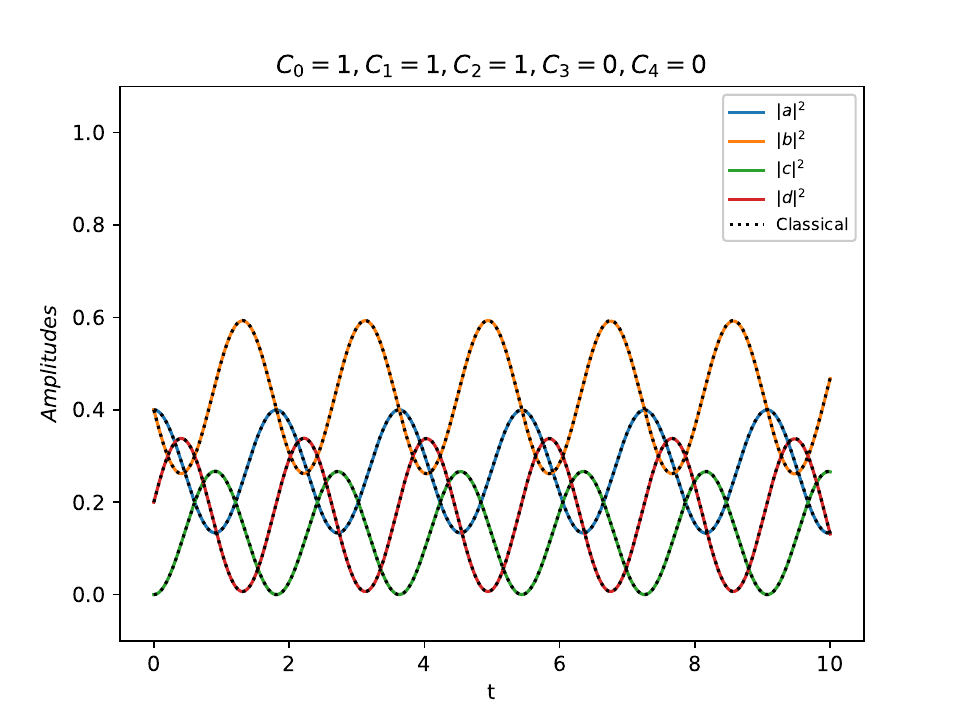}
	\end{subfigure}
	\hfill
	\begin{subfigure}[b]{0.49\textwidth}
		\includegraphics[width=\textwidth, height=7cm]{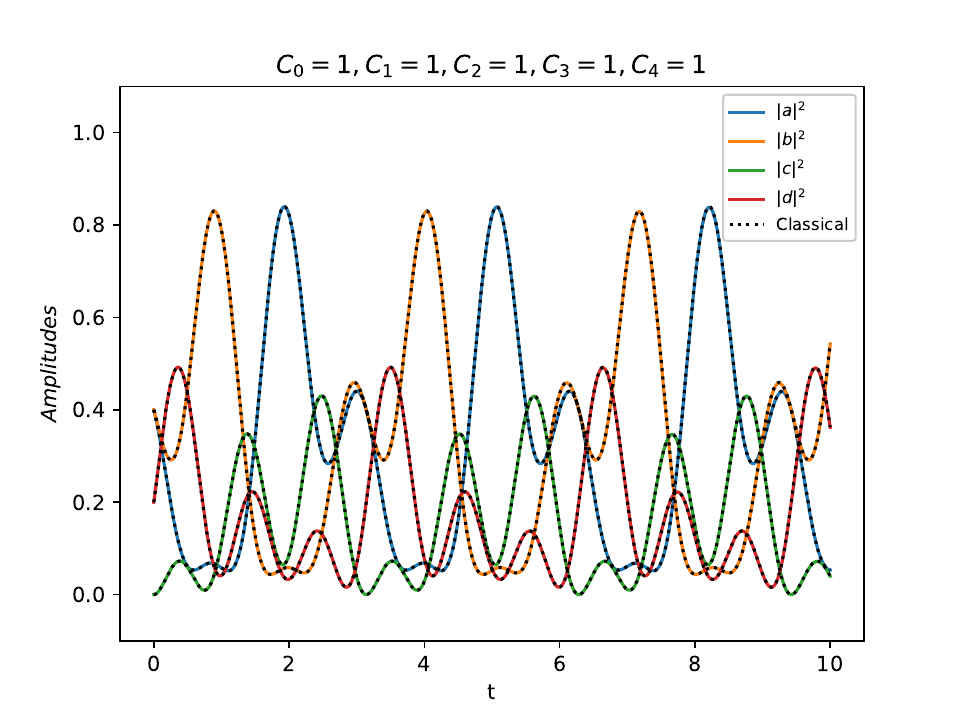}
	\end{subfigure}\caption{Time dependent amplitudes together with their classical counterparts, considering $a(0) = \sqrt{0.4}$ $, b(0) = \sqrt{0.4},$ $ c(0) = 0,$ $ d(0) = \sqrt{0.2}$. In the left panel, we analyze the dynamics driven by the non-entangling terms of the Hamiltonian, while in the right panel, we include the entangling terms.}
\label{amplitudes}
\end{figure}

As shown in Fig.~\eqref{amplitudes}, we analyze two different scenarios. In the left panel, we set the entangling terms in the Hamiltonian ($\hat{\sigma}_y \otimes \hat{\sigma}_y$ and $\hat{\sigma}_x \otimes \hat{\sigma}_y$) to zero. In contrast, in the right panel, we include these terms in the dynamics. This comparison is designed to demonstrate that the model is capable of correctly classicalize even the entangling components of the quantum Hamiltonian. As observed in both panels of Fig. \eqref{amplitudes}, the classical-like analogues 
shows exactly the same behaviour as quantum dynamics in all the presented cases.

The next quantity of interest for our analysis is the quaternionic population difference, which can be understood as an extension of the usual population difference of a two-level system to the four-level system case. If we define the quaternions $q_1$ and $q_2$ as
\begin{align}
    q_1 &= a + b i, & q_2 &= c + d i,
\end{align}
then the quaternionic population difference is defined as
\begin{equation}
    z(t) \equiv \abs{q_1}^2 - \abs{q_2}^2 = \abs{a}^2 + \abs{b}^2 - \abs{c}^2 - \abs{d}^2.
\end{equation}

Its classical counterpart, expressed in terms of the coordinates on $\mathbb{CP}^3$, is given by
\begin{equation}
    z(t) = \frac{\abs{x^0}^2 + \abs{x^1}^2 - \abs{x^2}^2 - 1}{\mathcal{N}}.
\end{equation}

\begin{figure}[H]
	\begin{subfigure}[b]{0.49\textwidth}
		\includegraphics[width=\textwidth, height=7cm]{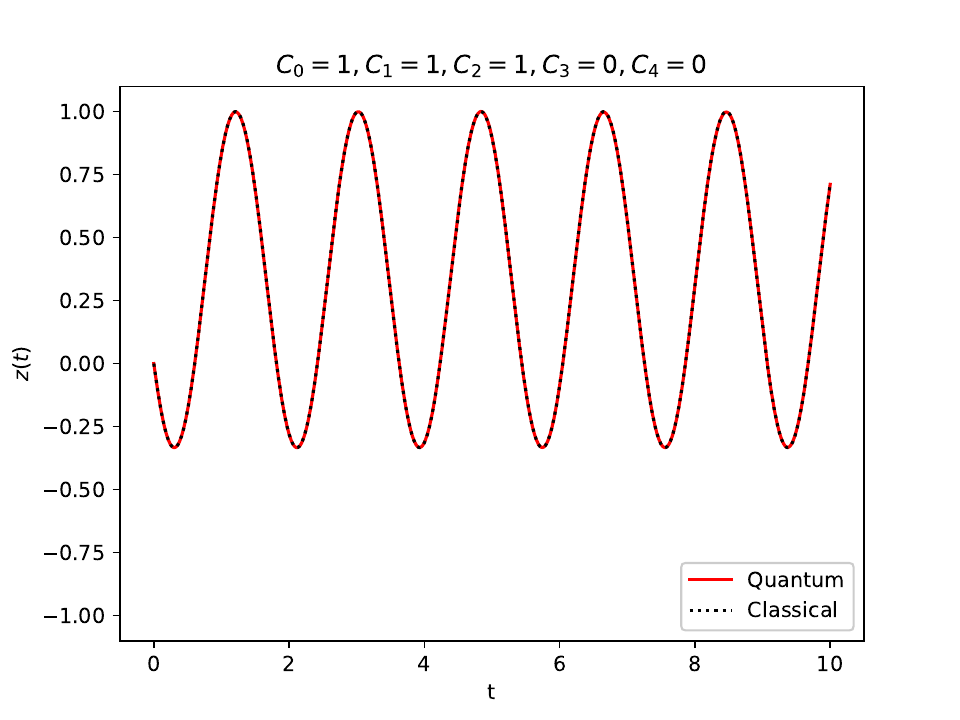}
	\end{subfigure}
	\hfill
	\begin{subfigure}[b]{0.49\textwidth}
		\includegraphics[width=\textwidth, height=7cm]{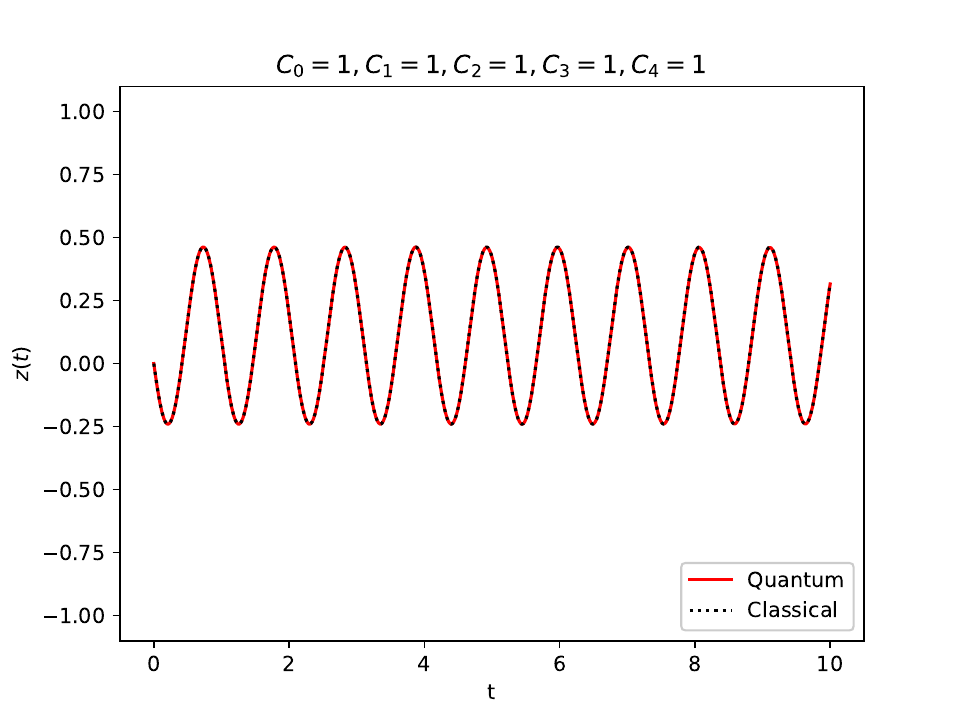}
	\end{subfigure}\caption{Time dependent quaternionic population difference, considering $a(0) = \sqrt{0.25},$ $ b(0) = \sqrt{0.25},$ $c(0) = \sqrt{0.25},$ $d(0) = \sqrt{0.25}$. The entangling terms of the Hamiltonian are only considered in the right panel.}\label{pop_diff}
\end{figure}

As shown in Fig. \eqref{pop_diff}, we once again consider two scenarios: in the left panel, the Hamiltonian excludes the entangling terms, while in the right panel, those terms are included. In both cases, we observe that the classical-like model exactly maps the quantum dynamics.

Now, we will compare the concurrence with its classical analogue in $\mathbb{CP}^3$. As mentioned earlier, the concurrence can be written as
\begin{equation}
    C(t) = 2\abs{ad - bc} = 2\frac{\abs{x^0 - x^1 x^2}}{\mathcal{N}}.
\end{equation}

\begin{figure}[H]
	\begin{subfigure}[b]{0.49\textwidth}
		\includegraphics[width=\textwidth, height=7cm]{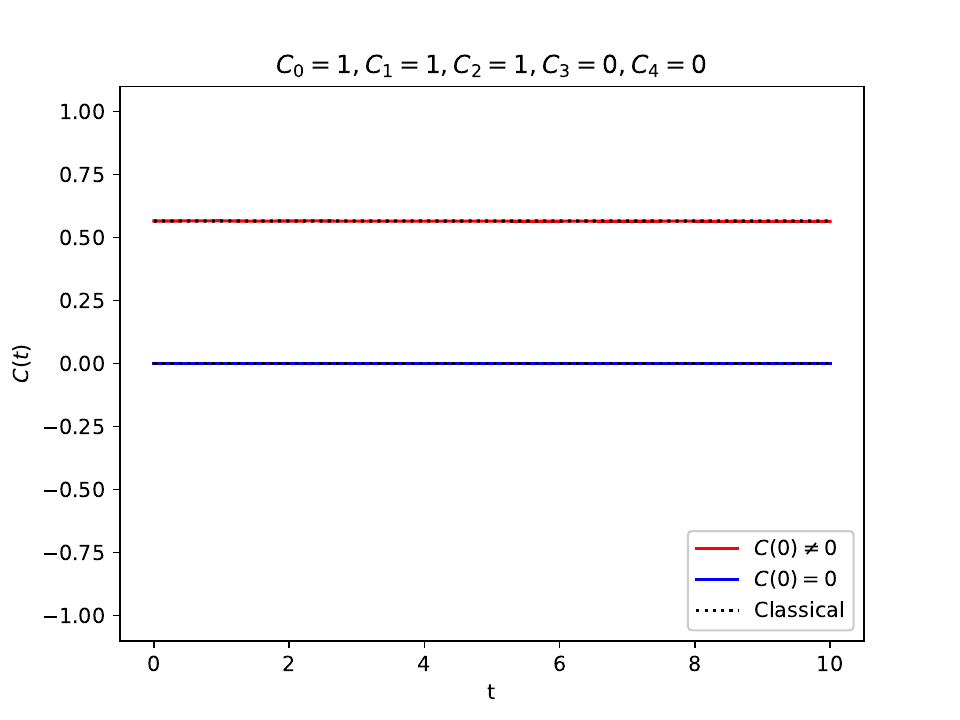}
	\end{subfigure}
	\hfill
	\begin{subfigure}[b]{0.49\textwidth}
		\includegraphics[width=\textwidth, height=7cm]{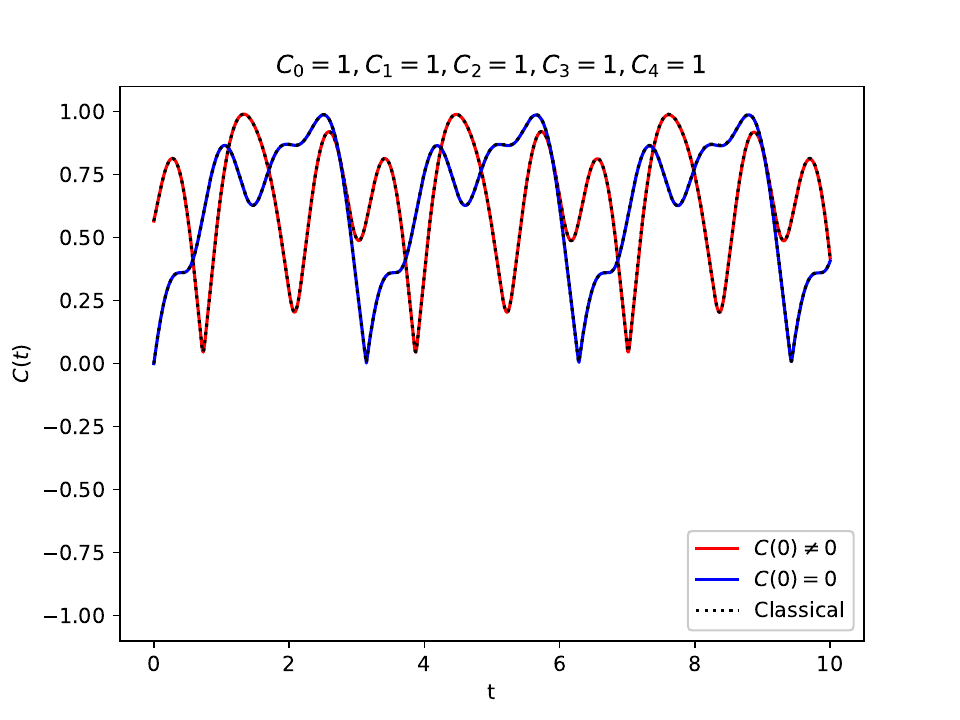}
	\end{subfigure}\caption{Time dependent concurrence. We have considered $a(0) = \sqrt{0.25},$ $ b(0) = \sqrt{0.25},$ $c(0) = \sqrt{0.25},$ $d(0) = \sqrt{0.25}$ for the blue line and $a(0) = \sqrt{0.4}$ $, b(0) = \sqrt{0.4},$ $ c(0) = 0,$ $ d(0) = \sqrt{0.2}$ for the red one. In the left panel, we analyze the dynamics considering the non-entangling terms of the Hamiltonian, while in the right panel, we include the entangling terms.
}\label{Fig_concurrence1}
\end{figure}

In Fig. \eqref{Fig_concurrence1}, we present two panels, each one displaying two different scenarios. In both panels, we consider initial states with either zero or non-zero concurrence. The left panel corresponds to the case where the entangling terms in the Hamiltonian are excluded, whereas the right panel includes all terms in the Hamiltonian. Once again, we observe that the classical-like analogue reproduces exactly the quantum dynamics.

\begin{figure}[H]
	\begin{subfigure}[b]{0.49\textwidth}
		\includegraphics[width=\textwidth, height=7cm]{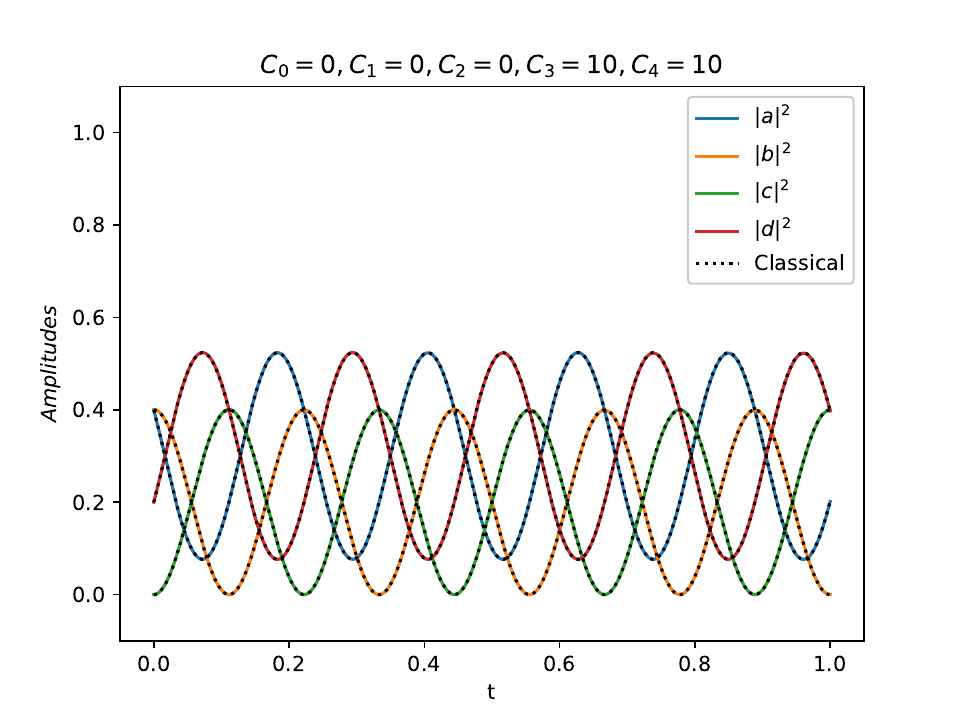}
	\end{subfigure}
	\hfill
	\begin{subfigure}[b]{0.49\textwidth}
		\includegraphics[width=\textwidth, height=7cm]{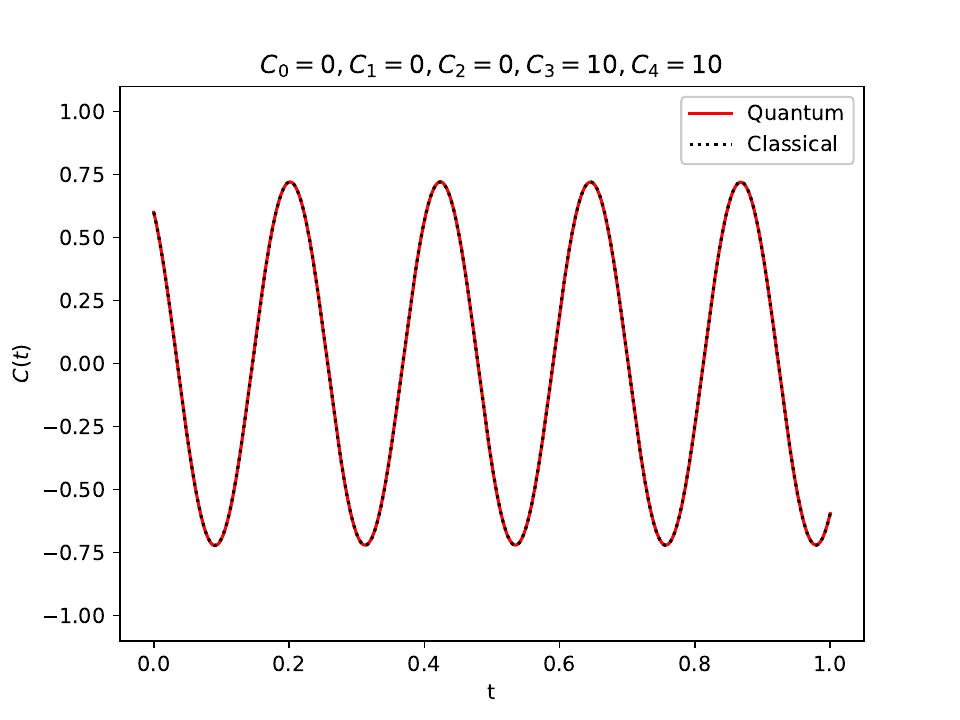}
	\end{subfigure}\caption{
    In the left panel, we plot the time dependent amplitudes, while in the right one, we consider the time dependet concurrence. In both panels, we have chosen as initial conditions $a(0) = \sqrt{0.4}$ $, b(0) = \sqrt{0.4},$ $ c(0) = 0,$ $ d(0) = \sqrt{0.2}$.}\label{fig4}
\end{figure}

Finally, in order to show how the model behaves under high entangling conditions, in Fig. \eqref{fig4}, we show the quantum amplitudes and the concurrence, considering $C_0 = C_1 = C_2 = 0$, $C_3 = C_4 = 10$, observing an increase of the oscillations' frequency. As we can see, even under high entangling conditions, the classical-like framework exactly reproduce the quantum behavior.

\section{Conclusions}\label{conclusions}
In this work, we have implemented a general and exact method for classicalizing any $N$-level quantum system, establishing a wide relationship between quantum and classical dynamics. By using the geometry of complex projective spaces ($\mathbb{CP}^{N-1}$), the method here employed systematically transforms quantum evolution into a set of $N-1$ Hamilton’s equations, fully preserving the behavior of the original system, including purely quantum phenomena such as entanglement. We have established a five steps method for classicalizing these systems, which are: defining coordinates in $\mathbb{CP}^{N-1}$, expressing the wave function in these coordinates, computing the classical Hamiltonian, deriving the symplectic form, and formulating Hamilton’s equations. This method is not only exact but also automatable, making it applicable to diverse systems without requiring huge case-specific adjustments.

In order to illustrate the power of the method, we have applied it to a paradigmatic system: two interacting qubits in $\mathbb{CP}^3$. This example demonstrates that the classical framework accurately reproduces key quantum observables, such as state populations, quaternionic population differences, and concurrence, even under Hamiltonian terms that generate entanglement.  This exact correspondence between classical trajectories and quantum evolution remarks the method’s ability to capture complex quantum effects without approximation.

The methodology can be highlighted by its generality and scalability. Unlike some mappings limited to specific Hamiltonians or certain regimes, the procedure here implemented is universal, applicable to any $N$-state system given its quantum Hamiltonian. The symplectic structure of $\mathbb{CP}^{N-1}$, derived from the Kähler potential, ensures that geometric properties scale predictably with dimension, facilitating extension to larger systems. 

This work could be extended in several ways. For example, the methodology can be applied to open quantum systems, incorporating dissipation and decoherence, where
a full system plus bath formalism can be studied (work in progress).
 Moreover, applying the method to larger systems, such as multi-qubit networks or higher-dimensional qudits, could yield computational advantages and a different and more geometric point of view.

In summary, we have implemented a unified framework that exactly classicalizes quantum systems, demonstrating its efficacy in the context of two-qubit entanglement and showing its general applicability to $N$-state systems. By combining mathematical rigor, computational efficiency, and geometric intuition, the method not only advances our theoretical understanding of the quantum-classical correspondence but also unlocks new possibilities for simulation across physics, chemistry, and beyond.

\section*{Acknowledgements}
D. M. -G. acknowledges Fundación Humanismo y Ciencia for financial support. D. M. -G. and P. B. acknowledge Generalitat Valenciana through PROMETEO PROJECT CIPROM/2022/13.

\bibliographystyle{unsrt}
\bibliography{referencias.bib}
\end{document}